\documentclass[12pt]{article}

  \newcounter{sectionc}\newcounter{subsectionc}\newcounter{subsubsectionc}
  \renewcommand{\section}[1] {\par\vspace*{0.6cm}\addtocounter{sectionc}{1} 
  \setcounter{subsectionc}{0}\setcounter{subsubsectionc}{0}\noindent 
  	{\normalsize\bf\thesectionc. #1}\par\vspace*{0.4cm}}
  \renewcommand{\subsection}[1] {\par\vspace*{0.6cm}\addtocounter{subsectionc}{1} 
  	\setcounter{subsubsectionc}{0}\noindent 
  	{\normalsize\it\thesectionc.\thesubsectionc. #1}\par\vspace*{0.4cm}}
  \renewcommand{\subsubsection}[1]
  {\vspace*{0.6cm}\addtocounter{subsubsectionc}{1}
  	\noindent {\normalsize\rm\thesectionc.\thesubsectionc.\thesubsubsectionc. 
  	#1}\par\vspace*{0.4cm}}

  \newcounter{appendixc}
  \newcounter{subappendixc}[appendixc]
  \newcounter{subsubappendixc}[subappendixc]

  \renewcommand{\appendix}[1] {\vspace*{0.6cm}
          \refstepcounter{appendixc}
          \setcounter{figure}{0}
          \setcounter{table}{0}
          \setcounter{equation}{0}
          \renewcommand{\thefigure}{\Alph{appendixc}.\arabic{figure}}
          \renewcommand{\thetable}{\Alph{appendixc}.\arabic{table}}
          \renewcommand{\theappendixc}{\Alph{appendixc}}
          \renewcommand{\theequation}{\Alph{appendixc}.\arabic{equation}}
          \noindent{\bf Appendix \theappendixc #1}\par\vspace*{0.4cm}}

\def\title#1{\begin{center}{\normalsize\bf #1}\end{center}\smallskip}
\def\author#1{\begin{center}{\footnotesize #1}\end{center}}
\def\address#1{\begin{center}{\footnotesize\it #1}\end{center}\smallskip}

  \def\abstract#1{{\vspace*{0.9 cm}
  	\centering{\begin{minipage}{12.2truecm}\footnotesize\baselineskip=12pt\noindent
  	\centerline{\footnotesize ABSTRACT}\vspace*{0.3cm}
  	\parindent=0pt #1
  	\end{minipage}}\par}} 

  \renewenvironment{thebibliography}[1]
        {\section{References}
  	\begin{list}{[\arabic{enumi}]}
  	{\usecounter{enumi}\setlength{\parsep}{0pt}
  \setlength{\leftmargin 1.25cm}{\rightmargin 0pt}
  	 \setlength{\itemsep}{0pt} \settowidth
  	{\labelwidth}{#1.}\sloppy}}{\end{list}}

  \newcounter{itemlistc}
  \newcounter{romanlistc}
  \newcounter{alphlistc}
  \newcounter{arabiclistc}

\long\def\@makecaption#1#2{
   \vskip 10pt 
   \setbox\@tempboxa\hbox{\footnotesize #1: #2}
 \ifdim \wd\@tempboxa >\hsize \footnotesize #1: #2\par \else \hbox
 to\hsize{\hfil\box\@tempboxa\hfil}  
   \fi}

  \def\@citex[#1]#2{\if@filesw\immediate\write\@auxout
  	{\string\citation{#2}}\fi
  \def\@citea{}\@cite{\@for\@citeb:=#2\do
  	{\@citea\def\@citea{,}\@ifundefined
  	{b@\@citeb}{{\bf ?}\@warning
  	{Citation `\@citeb' on page \thepage \space undefined}}
  	{\csname b@\@citeb\endcsname}}}{#1}}

   1
   1
   1

\newcommand{\alp}{\alpha}

\newcommand{\tht}{\theta}

\newcommand{\kp}{\kappa}
\newcommand{\lmd}{\lambda}
\newcommand{\Lmd}{\Lambda}

\newcommand{\Sgm}{\Sigma}
\newcommand{\vph}{\varphi}

\newcommand{\be}{\begin{equation}}
\newcommand{\ee}{\end{equation}}
\newcommand{\bea}{\begin{eqnarray}}
\newcommand{\eea}{\end{eqnarray}}
\newcommand{\eql}{\!\!\!&=\!\!\!&}

\newcommand{\tr}{{\rm tr}}

\newcommand{\diag}{{\rm diag}}
\newcommand{\der}{\partial}
\newcommand{\dr}{\!\!d}

\newcommand{\vev}[1]{\langle #1 \rangle}

\newcommand{\brkt}[1]{\left( #1 \right)}
\newcommand{\brc}[1]{\left\{ #1 \right\}}

\newcommand{\abs}[1]{\left| #1 \right|}

\renewcommand{\Im}{{\rm Im}}

\newcommand{\cN}{{\cal N}}
\newcommand{\cO}{{\cal O}}

\renewcommand{\ge}[2]{e_{#1}^{\;\;#2}}

\newcommand{\fG}{f_G}

\newcommand{\NP}[1]{{\it Nucl.~Phys.}~{\bf #1}}
\newcommand{\PL}[1]{{\it Phys.~Lett.}~{\bf #1}}

\newcommand{\PR}[1]{{\it Phys.~Rev.}~{\bf #1}}
\newcommand{\PRL}[1]{{\it Phys.~Rev.~Lett.}~{\bf #1}}
\newcommand{\PTP}[1]{{\it Prog.~Theor.~Phys.}~{\bf #1}}

\newcommand{\JH}[1]{{\it JHEP}~{\bf #1}}

\begin{document}
\title{
 Geometry mediated supersymmetry breaking\footnote{
Talk at the 12th International Conference on 
Supersymmetry and Unification of Fundamental Interactions, 
June 17-23, 2004, in Epochal Tsukuba, Tsukuba, Japan. }
}

\author{YUTAKA SAKAMURA} 

\address{
Department of Physics, Korea Advanced Institute of Science and Technology \\
373-1 Guseong-dong Yuseong-gu, Daejeon 305-701, Korea
\\ {\rm\normalsize E-mail: sakamura@muon.kaist.ac.kr}}

\abstract{We investigate SUSY breaking mediated through the deformation 
of the spacetime geometry due to the backreaction of a nontrivial configuration
of a bulk scalar field. 
To illustrate its features, we work with a toy model 
in which the bulk is four dimensions. 
Using the superconformal formulation of SUGRA, 
we provide a systematic method of deriving the 3D effective action 
expressed by the superfields. 
}

\normalsize\baselineskip=15pt

\section{Introduction}
When we construct brane-world models~\cite{ADD}, 
the size of the extra dimensions has to be stabilized. 
One of the main stabilization mechanisms is proposed in Ref.~\cite{GW}, 
and similar mechanisms have also been studied~\cite{dewolfe,EMS}. 
These mechanisms involve a bulk scalar field that has a nontrivial 
vacuum configuration. 
In such a case, the background geometry receives the backreaction 
of the scalar configuration. 
However, effects of such backreaction have been neglected in most works. 
Here, we investigate the SUSY breaking effects mediated through 
the deformation of the spacetime geometry due to 
the backreaction~\cite{geoSB4D}. 

In order to focus on the effects of SUSY breaking through 
the spacetime geometry, 
we consider a situation where a scalar field in the hidden sector 
has a non-BPS configuration. 
Then, the dominant contribution to SUSY breaking in the visible sector 
comes from the geometry-mediated effects. 
In order to understand qualitative features of this type of scenario, 
we work with a simplified toy model, 
in which the bulk spacetime is four dimensions and the effective theory 
is three-dimensional. 
An interesting example of the stabilization mechanisms is proposed 
in Ref.~\cite{EMS}. 
In this article, the authors found a non-BPS solution in the 4D SUGRA, 
which stabilizes the radius of the extra dimension and simultaneously 
generates a warped geometry. 

In the following, we will assume the existence of a non-BPS solution 
in 4D SUGRA, and derive the action on that background in terms of 
3D superfields. 
Our method of deriving the action is based on the superconformal formulation 
of SUGRA~\cite{KU}, but is easy to handle thanks to 
the superfield formalism.  

\section{Invariant action}
Here, we choose the direction of $y\equiv x_2$ as the extra dimension, 
and $m,n=0,1,3$ denote the 3D vector indices. 
The underbarred indices denote the local Lorentz indices. 
We will consider the following gravitational background. 
\bea
 \vev{\ge{\mu}{a}} \eql \diag(e^A,e^A,1,e^A), \;\;\;\;\;
 \vev{\psi_\mu^\alp}=0, \nonumber\\
 \vev{A_m} \eql 0,  \;\;\;\;\;
 \vev{A_y}=\frac{\kp}{2}\Im(\bar{\phi}^{\rm h}_{\rm bg}
 \der_y\phi^{\rm h}_{\rm bg}), 
\eea
where $A(y)$ is a warp factor, 
$A_\mu$ is a gauge field of $U(1)_A$, a constant~$\kp\equiv 1/M_{\rm pl}$ 
is the gravitational coupling, 
and $\phi^{\rm h}_{\rm bg}(y)$ is the background scalar configuration 
in the hidden sector. 
On the above background, an superconformal invariant action can be expressed 
in terms of 3D superfields as follows. 
\bea
 S_{\rm bg} \eql \int\dr^4x d^2\tht\;\left[
 2e^AG_{I\bar{J}}D^\alp\bar{\vph}^{\bar{J}}D_\alp\vph^I 
 +4e^{2A}\Im\brc{G_I\nabla_y\vph^I+e^{2i\tht_A}W(\vph)} \right. \nonumber\\
 &&\hspace{20mm}\left.+\frac{8}{3}e^{2A}\vev{A_y}G
 +\frac{e^{2A}}{2}\tr\brc{T^{-1}(u^g)^2-T(w^g)^2}\right], \label{Sinv}
\eea
where $G(\bar{\vph},\vph)$ is a real function of $\vph^I$ and 
$\bar{\vph}^{\bar{J}}$, 
and $W(\vph)$ is a holomorphic function of $\vph^I$. 
3D scalar superfields~$\vph^I$ is constructed 
from 4D chiral multiplets, and the 3D spinor superfields~$u^g_\alp$ 
and $w^g_\alp$ are constructed from a 4D vector multiplet. 
Especially, $w^g_\alp$ corresponds to the 3D superfield strength. 
Derivative operators~$D_\alp$ and $\nabla_y$ are covariant derivatives 
for the gauge symmetry. 
(For their explicit definitions, see Ref.~\cite{geoSB4D}.)
Furthermore, 
\be
 T\equiv 1+e^A\tht\vev{A_y}
\ee
is a spurion superfield corresponding to the radion superfield.

\section{Action after gauge fixing}
In order to obtain the Poincar\'{e} SUGRA, we have to remove the redundant 
superconformal gauge symmetry by imposing the gauge fixing conditions. 
After the gauge fixing, the invariant action becomes 
\bea
 S_{\rm vis} \eql \int\dr^4xd^2\tht\;\left[
 e^A\sum_iD^\alp\bar{\vph}^{\bar{i}}D_\alp\vph^i
 +4e^{2A}\Im\brc{\frac{1}{2}\sum_i\bar{\vph}^{\bar{i}}\nabla_y\vph^i
 +e^{(\kp^2/2)\abs{\phi_{\rm bg}^{\rm h}}^2}P_{\rm vis}(\vph)}\right.
 \nonumber\\
 &&\hspace{20mm}\left.
 +\frac{4}{3}e^{2A}\vev{A_y}\sum_i\bar{\vph}^{\bar{i}}\vph^i
 +\frac{e^{2A}}{2}\tr\brc{(u^g)^2-(w^g)^2}\right]
 \nonumber\\
 &&+\kp^2\int\dr^4x\;e^{3A}\left[
 \Im\brc{-2\sum_if^G_{\rm bg}\phi^i\bar{f}^{\bar{i}}
 +6\hat{f}^\Sgm_{\rm bg}P_{\rm vis}
 -\frac{4\vev{A_y}}{3\kp^2}\brkt{3P_{\rm vis}-\sum_i\phi^i
 \frac{\der P_{\rm vis}}{\der \phi^i}}}\right. \nonumber\\
 &&\hspace{27mm}\left.
 -\frac{\vev{A_y}}{2\kp^2}\tr\brc{(\lmd_1^g)^2+(\lmd_2^g)^2}
 +\cdots\right]+\cO(\kp^4), 
\eea
where the scalar~$\phi^i$, $\lmd_1^g$, and $\lmd_2^g$ are 
the lowest components of $\vph^i$, $u^g$, and $w^g$, respectively. 
Here, we have assumed the minimal K\"{a}hler potential, 
and $P$ is a superpotential which is related to $W$ in Eq.(\ref{Sinv}) as
\be
 W=\brkt{\frac{2}{3}}^{3/2}\kp^3\brkt{\vph^\Sgm}^3
 \brc{P_{\rm hid}(\vph^{\rm h})+P_{\rm vis}(\vph^i)},  
\ee 
where $\vph^\Sgm$ is a compensator superfield, $\vph^{\rm h}$ is 
a hidden-sector superfield whose lowest component has a nontrivial vacuum 
configuration, and $\vph^i$ are matter superfields in the visible sector. 
Since we are interested only in the visible sector, 
we have dropped the fluctuation modes of $\vph^{\rm h}$ in the hidden sector. 
Note that the action is supersymmetric at $\cO(\kp^0)$. 
This is a trivial result. 
Because all the SUSY breaking effects in the visible sector 
are induced through the deformation of the geometry, they will vanish 
in the limit that the gravity is turned off. 
Thus, SUSY breaking terms appear at $\cO(\kp^2)$. 
$f^G_{\rm bg}(y)$ and $\hat{f}^\Sgm_{\rm bg}(y)$ are some functions of 
the warp factor~$A(y)$ and the scalar background~$\phi^{\rm h}_{\rm bg}(y)$, 
which will vanish in the BPS limit.

\section{Summary and comments}
We have discussed the effects of SUSY breaking mediated by 
the deformation of the spacetime geometry due to the backreaction 
of a bulk scalar field. 
We derived the action expressed by 3D superfields and 
the SUSY breaking terms. 

Scales introduced in the theory are the 4D Planck mass~$M_{\rm pl}$, 
the mass parameters for the matter fields~$m_i$, the characteristic scale 
of the hidden sector dynamics~$\Lmd$, and the compactification 
scale~$r^{-1}$. 
In terms of these scales, the SUSY-breaking scalar masses induced 
in the visible sector have a form of 
\be
 m_S^2=\frac{\Lmd^3 m_i}{M_{\rm pl}^2}\alp(m_i,\Lmd,r), 
\ee
where $\alp(m_i,\Lmd,r)$ is a dimensionless function expressed by 
an overlap integral of the mode functions and 
the functions~$f^{\rm h}_{\rm bg}$, $f^G_{\rm bg}$, and $\vev{A_y}$. 
The gaugino mass is induced by the nonzero $\vev{A_y}$, 
and roughly estimated as 
\be
 m_g\sim\frac{\Lmd^3}{M_{\rm pl}^2}. 
\ee
Here, we have supposed that $\arg(\phi^{\rm h}_{\rm bg})$ varies with 
$\cO(1)$ amplitude as $y$ goes from $0$ to $\pi r$. 

Since all the SUSY breaking effects discussed here 
are suppressed by the Planck mass, 
our scenario can be considered as a kind of the gravity mediation. 
However, there are some points that should be noticed. 
First, from the viewpoint of the effective theory, 
SUSY breaking discussed here cannot be regarded as a spontaneous breaking 
because the order parameter of SUSY breaking is roughly 
of $\cO(\Lmd)$ 
and is generally higher than the compactification scale~$r^{-1}$, 
which is the cut-off scale of the 3D effective theory. 
Second, the induced SUSY breaking scale in the effective theory 
can be suppressed by the overlap integral of 
$\hat{f}^\Sgm_{\rm bg}(y)$, $\fG(y)$, and the mode functions.  
Third, the gaugino mass can be induced by non-zero $\vev{A_y}$ 
without introducing a non-minimal gauge kinetic function. 
This is very similar to the Scherk-Schwarz (SS) SUSY breaking~\cite{SS} 
interpreted in the Hosotani basis. 
However, this breaking is irrelevant to the $U(1)_R$ twisting 
since $U(1)_R$ is a symmetry {\it after} the gauge fixing 
and independent of $U(1)_A$, which is completely fixed 
by the gauge fixing condition.  
In addition, $\vev{A_y}$ is not an input parameter as in the SS breaking, 
but is determined by the bulk scalar dynamics. 
Further, the non-zero $\vev{A_y}$ indicates that the background configuration 
is non-BPS. 
Thus, it inevitably leads to SUSY breaking terms that associate with 
$\hat{f}^\Sgm_{\rm bg}$ and $\fG$, 
to which the SS breaking does not have any resemblances. 

To investigate more phenomenological aspects, we should extend our discussion 
to 5D SUGRA. 
Note that the procedure explained here requires only the knowledge of 
the superconformal formulation of 4D SUGRA and the 3D superfield formalism. 
The 4D superfield formalism is not necessary. 
Therefore, the extension to 5D SUGRA only requires the 5D superconformal 
formulation~\cite{KO} and the well-known 4D $\cN=1$ superfield formalism. 
This work is done in Ref.~\cite{AS}.

\section{Acknowledgements}
This work was supported by the Astrophysical Research Center 
for the Structure and Evolution of the Cosmos (ARCSEC) 
funded by the Korea Science and Engineering Foundation 
and the Korean Ministry of Science. 

\bibliographystyle{plain}

\end{document}